\documentclass[aps,prb,twocolumn,groupedaddress]{revtex4}

\usepackage{graphicx}

\begin{document}

\title{Vibrational and electronic properties of the B$_7$Al$_2$ cluster}

\author{P. L. Rodr\'iguez-Kessler}
\email{plkessler@cio.mx}
\affiliation{Centro de Investigaciones en \'Optica A.C., Loma del Bosque 115, Lomas del Campestre, Leon, 37150, Guanajuato, Mexico}

\date{\today}

\begin{abstract}
In this work, we employ density functional theory (DFT) to explore the structure of boron clusters doped with two aluminium atoms (B$_7$Al$_2$ or Al$_2$B$_7$). The results show that the most stable structure is a bipyramidal configuration formed by a B$_7$ ring coordinated with two Al atoms, while the higher energy isomers correspond to peripheral Al$_2$-doped B$_7$ structures. The IR spectra and density of states reveal remarkable differences between the global minimum and the higher energy isomers. 

\end{abstract}


\maketitle

\section{Introduction}

Over the past few decades, significant theoretical and experimental research has been conducted on boron clusters. These clusters exhibit unique structural and electronic characteristics, along with unusual chemical bonding, which have garnered considerable interest in the scientific community.\cite{https://doi.org/10.1002/anie.200701915,https://doi.org/10.1002/anie.200903246} Boron clusters are extensively researched in materials science and nanotechnology due to their electron-deficient nature. This electron deficiency allows them to form a variety of stable and metastable structures, making them versatile building blocks for various applications. Doping these clusters with other elements can significantly modify their electronic, magnetic, and chemical properties, paving the way for numerous potential applications.\cite{D4CP00296B} Although substantial progress has been made in recent years in the experimental and theoretical studies of transition-metal-doped boron clusters, research on boron clusters doped with two transition metal atoms remains relatively limited.\cite{molecules28124721} In 2006, Zhai et al. studied the geometric structure and chemical bonding of a binary Au$_2$B$_7^-$ clusters.\cite{doi:10.1021/jp0559074} In 2014 Li and coworkers revealed that Ta$_2$B$_6^{-/0}$ clusters adopt a bipyramidal structure, with a B$_6$ ring sandwiched between two Ta atoms.\cite{https://doi.org/10.1002/anie.201309469} Wang and colleagues studied lanthanide-doped boron clusters, Ln$_2$B$_n$ (n=7, 9) and discovered inverse sandwich motifs within this series of chemical elements.\cite{doi:10.1073/pnas.1806476115} In particular, studies on M$_2$-doped boron clusters M$_{2}$B$_7$ involving alkaline earth metals (M = Be, Mg, Ca) as well as first-row transition metals (M=Sc, V, Fe, Co, Ni) also showed similar structures including enhanced reactivities.\cite{JIA2014128,PHAM2019186,LI202325821,OLALDELOPEZ2024} Recently, Yue et al., investigated the fluxional and dynamic behavior of the Al$_2$B$_8$ cluster. Interestingly they found that the Al$_2$B$_8$ cluster adopt a simple sandwich structure with a heptacoordinate B$_8$ molecular wheel. The Al$_2$ component is divided into two isolated Al atoms, which are situated at above and below of the B$_8$ wheel.\cite{D2RA07268H} Herein, we evaluate the case of Al$_2$B$_7$ cluster in order to identify if the removal of one B atom in Al$_2$B$_8$ makes the cluster more rigid, according to the previous reports on M$_2$B$_7$ clusters. The structural, vibrational, and electronic properties are discussed. The data provided in this preprint serves for further investigation into specific potential applications for the clusters. 

\section{Computational Details}

Calculations performed in this work are carried out by using density functional theory (DFT) as implemented in the Orca quantum chemistry package. \cite{10.1063/5.0004608}. The Exchange and correlation energies is addressed by the PBE0 functional in conjunction with the Def2TZVP basis set.\cite{10.1063/1.478522,B508541A} Atomic positions are self-consistently relaxed through a Quasi-Newton method employing the BFGS algorithm. The SCF convergence criteria for geometry optimizations are achieved when the total energy difference is smaller than 10$^{-8}$ au, by using the TightSCF keyword in the input. The  Van  der  Waals  interactions  are  included in the exchange-correlation functionals with empirical dispersion corrections of Grimme DFT-D3(BJ). The total density of states (DOS) and partial density of states (PDOS) for clusters and complexes were obtained by using the Multiwfn program.\cite{https://doi.org/10.1002/jcc.22885} The IR spectra is obtained by standard diagonalization of the Hessian matrix by using the NumFreq keyword in the input.

\section{Results}

\begin{figure}[h!]
\begin{center}
\scriptsize
 \includegraphics[scale=0.35]{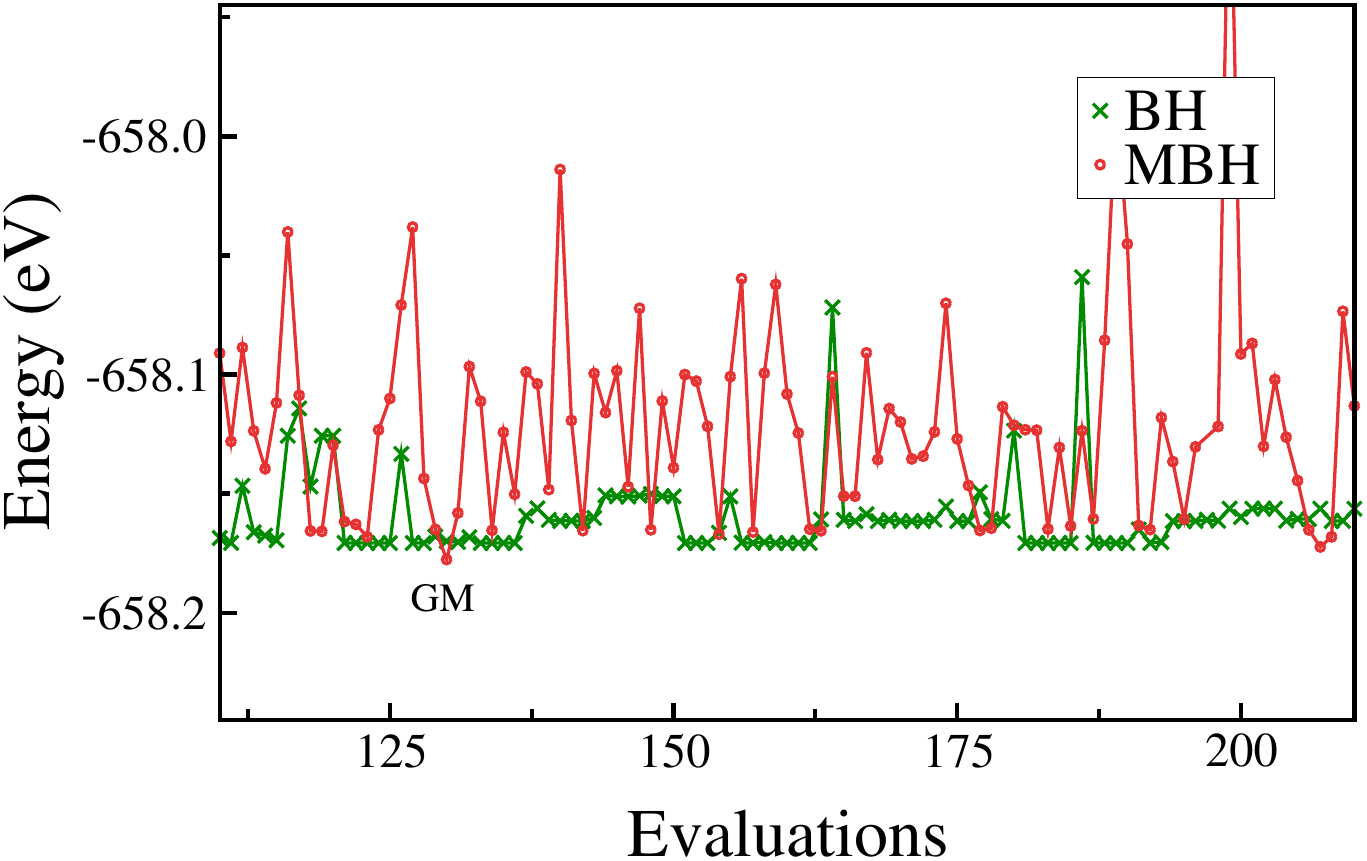}
	\caption{\label{figure_BH}Potential energy surface for standard basing hopping (BH) and modified basing hopping (MBH) for B$_7$Al$_2$ clusters.}
\end{center}
\end{figure}

The structures of the B$_7$Al$_2$ clusters are obtained using a modified basin-hoping (MBH) structure search method over seven generations, as implemented in previous works.\cite{D0CP06179D,OLALDELOPEZ2024} Interestingly, the standard basin-hopping method encounters difficulties in locating the global minimum (GM) structure of the B$_7$Al$_2$ cluster. This is because the Al atoms tend to remain together, and the random perturbations cannot efficiently explore all possible combinations in the binary alloy. This can be illustrated with the potential energy surface, in which the standard BH method get stuck in local minimum configurations (Figure~\ref{figure_BH}). In literature, the BH method has been improved by adding local and non-local operations for more efficient exploration of the potential energy surface.\cite{doi:10.1021/ci400224z} Here we have used a modified basin hopping method (MBH) which includes random exchanges of atomic species along with standard random perturbations. The efficiency of the MBH method will be tested by considering several classes of systems, such as LJ clusters, metal clusters and alloys,\cite{RODRIGUEZKESSLER2024122376} which will be reported in an upcoming manuscript.

\begin{figure}[h!]
  \begin{tabular}{c}
      \includegraphics[scale=0.7]{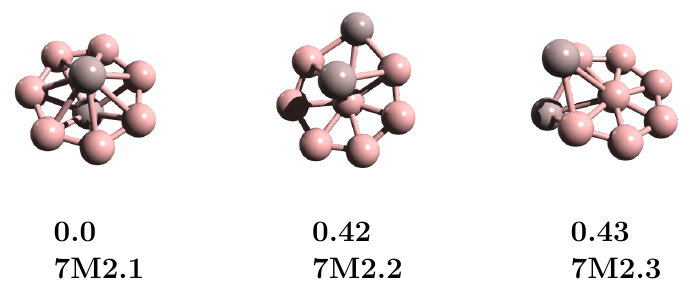} \\
      \includegraphics[scale=0.7]{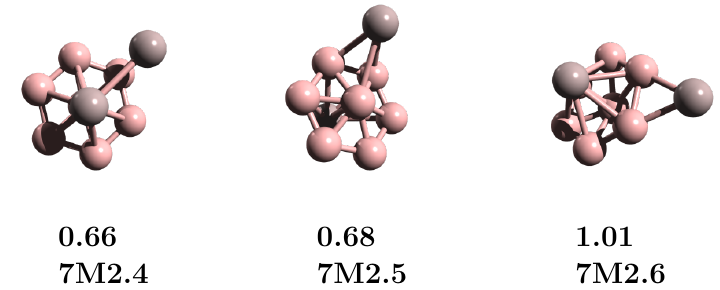} \\
  \end{tabular}
       \caption{\label{figure_struc}Lowest energy structures for B$_7$Al$_{2}$ clusters. For simplicity, the isomers are denoted by {\bf 7M2.y (y=1-6)}.\cite{rodríguezkessler2024structuresstabilitiesb7cr2clusters} For each structure, the relative energy (in eV) and isomer label are given. }
\end{figure}

\begin{figure}[h!]
\begin{center}
\scriptsize
 \includegraphics[scale=0.35]{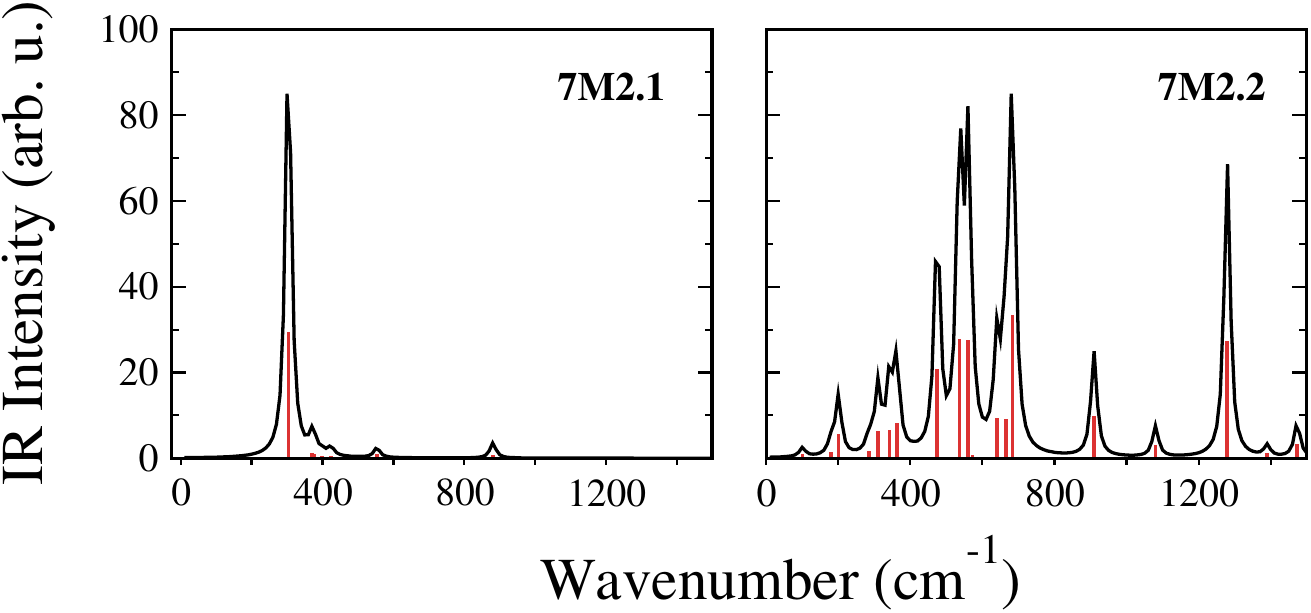}
	\caption{\label{figure_IR}IR spectra for {\bf 7M2.1} and {\bf 7M2.2} clusters obtained at the PBE0/Def2-TZVP level.}
\end{center}
\end{figure}

The results show that the lowest energy structure for B$_7$Al$_{2}$ cluster is an inverse sandwich structure ({\bf 7M2.1}), which is in good agreement with similar reports on B$_7$M$_{2}$ clusters.\cite{ZHANG2015131,HAO20181,JIA2014128} The next isomer found is a peripheral Al$_2$-doped B$_7$ structure ({\bf 7M2.2}) with a relative energy of 0.42 eV (Figure~\ref{figure_struc}). The average Al-B bond distance amounts 2.52 \AA, which is larger than the M-B bonds of B$_7$M$_{2}$ (M=Cr, Fe, Co, Ni) clusters, amounting 2.17-2.29 \AA.\cite{OLALDELOPEZ2024,rodríguezkessler2024structuresstabilitiesb7cr2clusters} The infrared (IR) spectrum of B$_7$Al$_{2}$ clusters is calculated for future experimental comparison.\cite{Wen2020,rodríguezkessler2024structuresstabilitiesb7cr2clusters,rodríguezkessler2024revisitingglobalminimumstructure} For {\bf 7M1.1} a single peak is found at 303.89 cm$^{-1}$, while for {\bf 7M1.2} three main peaks were found at 536.22, 638.31 and 1279.19 cm$^{-1}$ (Figure~\ref{figure_IR}). The lowest and highest vibrational frequencies for {\bf 7M1.1} are 303.89-881.34 cm$^{-1}$, while for {\bf 7M1.2} are 100.65-1473.28 cm$^{-1}$, respectively, suggesting more dispersion in {\bf 7M1.2} due to a less symmetrical structure.

\begin{figure}[h!]
 \resizebox*{0.50\textwidth}{!}{\includegraphics{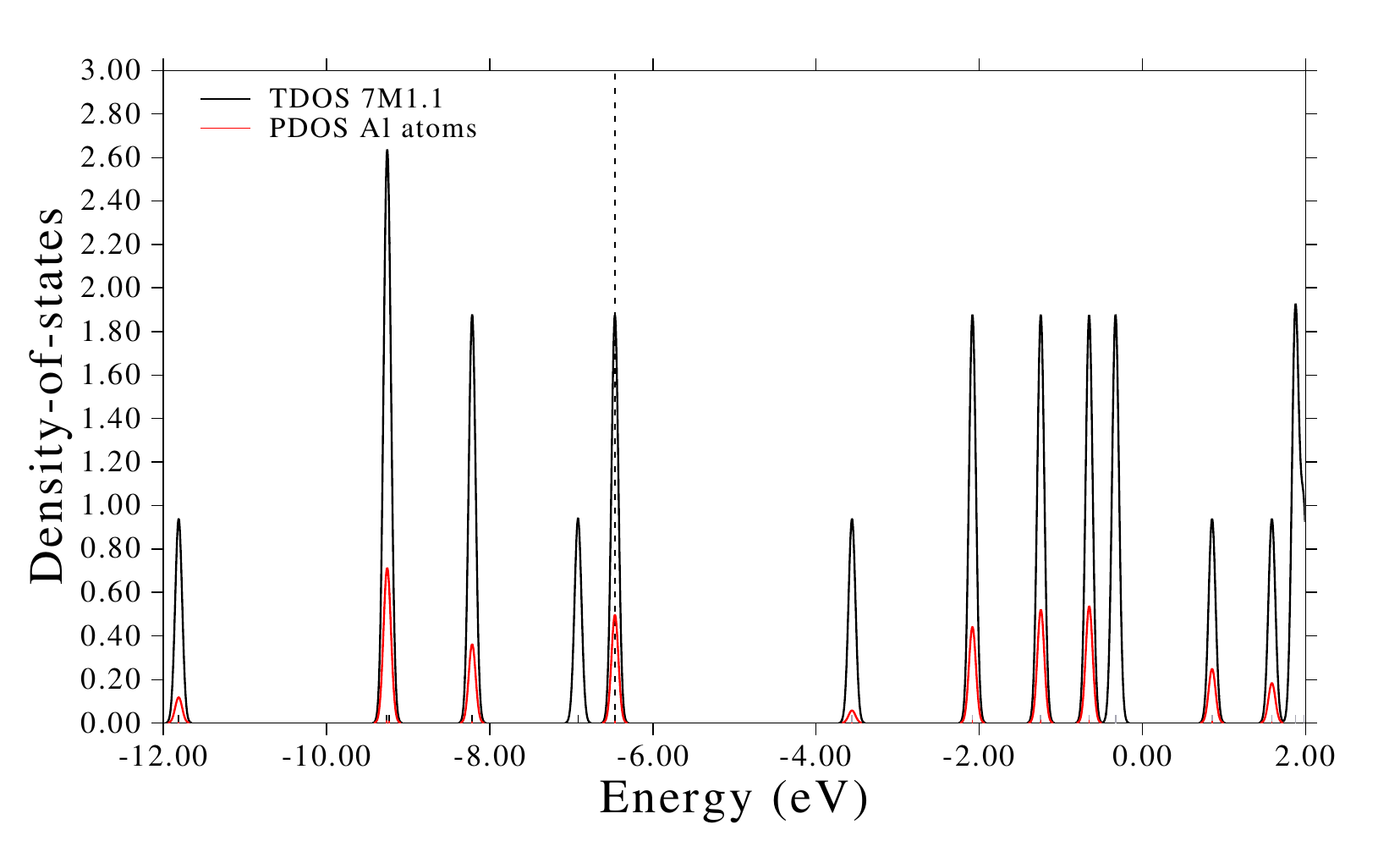}} 
\resizebox*{0.50\textwidth}{!}{\includegraphics{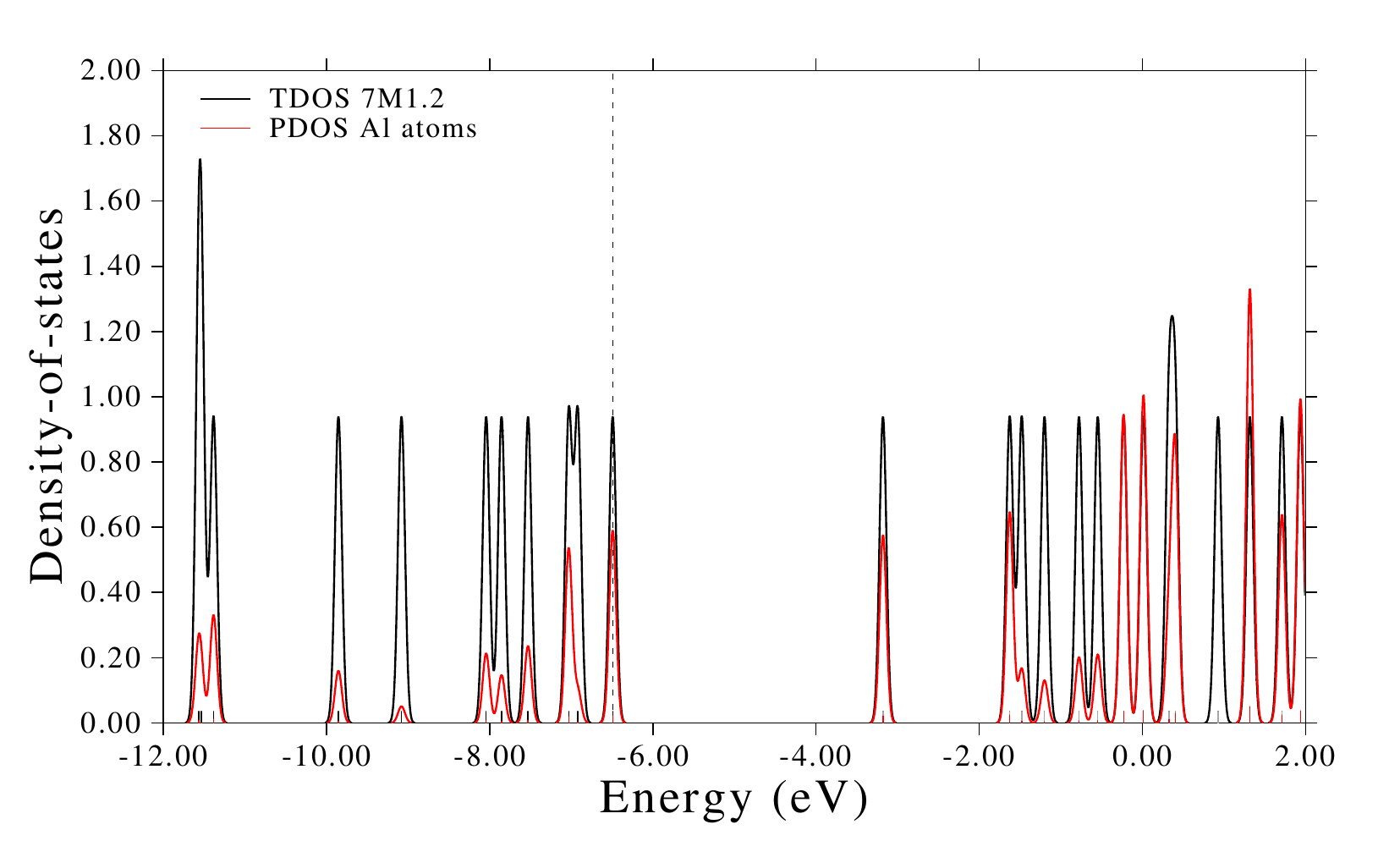}} 
\caption{\label{figura4}The total and partial density of states (TDOS, PDOS) {\bf 7M2.1} and {\bf 7M2.2} cluster. Vertical dashed line corresponds to the HOMO energy level.}
\end{figure}

To get insight into the electronic properties of the clusters, the total density of states (TDOS) for {\bf 7M2.1} and {\bf 7M2.2} cluster is depicted in Figure~\ref{figura4}. For {\bf 7M2.1}, the PDOS show more localized states, on the other hand for {\bf 7M2.2}, less localization with higher HOMO-LUMO gap, suggesting that the former is more reactive. This preliminary results help for further exploring the reactivity of the clusters, with possible applications in catalysis.\cite{OLALDELOPEZ2024,RODRIGUEZKESSLER2024141588,RODRIGUEZCARRERA2024122301,RODRIGUEZKESSLER201820636,doi:10.1021/acs.jpcc.9b03637,https://doi.org/10.1002/adts.202100043,D2CP05188E,10.1063/1.4935566,RODRIGUEZKESSLER201534,RODRIGUEZKESSLER201532,RODRIGUEZKESSLER2020155897} Moreover, the data provided in this preprint serves to enrich the libraries of M$_2$-doped B$_7$ clusters.

\section{Conclusions}
 In this work, we have employed density functional theory (DFT) to explore the structure of boron-aluminium clusters with the formula B$_7$Al$_2$. The results showed that the most stable structure is a bipyramidal configuration formed by a B$_7$ ring coordinated with two Al atoms, while the higher energy isomers correspond to peripheral Al$_2$-doped B$_7$ structures. The IR spectra and density of states revealed remarkable differences between the global minimum and the higher energy isomers.

\section{Acknowledgments}
P.L.R.-K. would like to thank the support of CIMAT Supercomputing Laboratories of Guanajuato and Puerto Interior.



\nocite{*}
\bibliographystyle{ieeetr}  
\bibliography{mendelei}

\end{document}